\documentclass[12pt]{iopart}

\usepackage{graphicx}
\usepackage{amssymb}

\begin{document}
\title{Study of the derivative expansions for the nuclear structure functions}

\author{I Ruiz Sim\'o and M J Vicente Vacas}
\address{Departamento de
F\'\i sica Te\'orica and IFIC, Centro Mixto Universidad de
Valencia-CSIC, Institutos de Investigaci\'on de Paterna,
Aptdo. 22085, E-46071 Valencia, Spain.}

\begin{abstract}
We study the convergence of the series expansions sometimes used in the analysis of the nuclear effects
in Deep Inelastic Scattering (DIS) proccesses induced by leptons. The recent advances in statistics and quality of the data, in particular for neutrinos calls for a good control of the theoretical uncertainties of the models used in the analysis.
Using realistic nuclear spectral functions which include nucleon correlations,
we  find that the convergence of the derivative expansions  to the full results is poor except at very low values of $x$.

\end{abstract}

\pacs{13.15.+g,13.60.Hb,25.30.-c}



\section{Introduction}

The advent of new and high statistics deeply inelastic neutrino scattering experiments 
apart from providing valuable data on $F_2$ and $x\,F_3$\cite{Conrad:1997ne, Tzanov:2005kr}
has shown again the importance of nuclear effects that render difficult the extraction of the nucleon 
structure functions. Furthermore, the use of nuclear targets is necessary due to the smallness of the cross sections. 
Qualitatively, the nuclear effects are well known. Shadowing, antishadowing, Fermi motion, binding,... have all been
widely studied for charged leptons in the context of the European Muon Collaboration (EMC) effect. For a review, 
see \cite{Geesaman:1995yd,Armesto:2006ph}. Moreover, it is well known that  nuclear effects are substantially different
for neutrino reactions due to the presence of the axial current and the different valence and sea quark contributions
for each observable~\cite{Schienbein:2007fs}. This situation asks for a
detailed and quantitative microscopical understanding of the nuclear effects, rather than the parametrizations that
have been used sometimes, like recently by the NuTeV Collaboration~\cite{Tzanov:2005kr}.

One of the basic ingredients in all calculations  is the nuclear spectral function. This presents some serious difficulties, 
as these functions are not so well known to the precision level reached by current experiments. Therefore, the  analysis might introduce unwanted model dependences. However, it was soon noticed that under some approximations
the nuclear structure functions could be written as  simple expansions on the  nucleon structure functions and their derivatives. All the nuclear information would then be encoded in
the expected values of some nuclear magnitudes, like the average kinetic energy of the nucleons or the mean nucleon removal energy. 
These kind of approximations, both for charged lepton and for neutrino induced reactions, have been widely used in the literature
\cite{Frankfurt:1985ui,CiofiDegliAtti:1989eg,CiofidegliAtti:1990dh,CiofidegliAtti:1991ae,Kulagin:1997vv,Miller:2001tg,Smith:2002ci,Ciofi degli Atti:2007vx}. The aim of this paper is to investigate the quality of these expansions and to what extent they could be used in the analysis of lepton deep inelastic scattering experiments. 
In order to do that, once the formalism is established in the next section, we will make a comparative study
for $F_{2,3}$ and for a few typical nuclear spectral functions used in the literature.

\section{Formalism}
\label{sec:Fo}
The nuclear structure functions can be written as a convolution of the nuclear spectral functions and nucleon structure functions. See for instance  \cite{Kulagin:2007ju} and references therein. In the rest frame of the nucleus, the $F_2$ and $F_3$ structure functions are 
\begin{eqnarray}
\fl
F_2^A(x,Q^2)&=&\sum_{\tau=p,n}\int\frac{d\epsilon\, d^3p}{(2\pi)^4}\mathcal{P}^\tau(\epsilon,\mathbf{p})\frac{(1+\gamma\frac{p_z}{M})}{\gamma^2}\left( 1+4\frac{p^2x'^2}{Q^2}+6\frac{x'^2\mathbf{p}^2_\bot}{Q^2}\right) F_2^\tau(x',Q^2)\label{eq:1},\\
\fl
F_3^A(x,Q^2)&=&\sum_{\tau=p,n}\int\frac{d\epsilon \,d^3p}{(2\pi)^4}\mathcal{P}^\tau(\epsilon,\mathbf{p})\left( 1+\frac{p_z}{\gamma M}\right) \frac{x'}{x} F_3^\tau(x',Q^2)\label{eq:2},
\end{eqnarray}
where $\mathcal{P}^{p(n)}(\epsilon,\mathbf{p})$ is the nuclear spectral function, normalized to the number of protons (neutrons) in the nucleus, and describes the probability of finding a proton (neutron) with momentum $\mathbf{p}$ and removal energy $\epsilon$. The four-momentum of the nucleon can be written as $p=(M+\epsilon,\mathbf{p})$, with $\epsilon\leq0$. The \textit{z} axis is oriented in such a way that $\mathbf{q}$ lies on it, $\mathbf{p}_\bot$ is the transverse momentum of the nucleon and $\gamma=\vert\mathbf{q}\vert/q^0$. Here,
$x'$ is the natural Bjorken variable for the nucleon in the nuclear medium, i.e. $x'=Q^2/(2p\cdot q)$; while $x$ is the Bjorken variable in the nucleon rest frame, $x=Q^2/(2Mq^0)$. They are related  by
\begin{equation}
x'=\frac{x}{z} \;\; \textrm{where}\;\; z=1+\frac{\epsilon}{M}+\gamma\frac{p_z}{M}.\label{eq:3}
\end{equation}

For isoscalar nuclei such as $^{40}Ca$, only the isoscalar component of the spectral function and the structure function have to be accounted for. 
When this is done  (\ref{eq:1}) and (\ref{eq:2}), these read as
\begin{eqnarray}
\fl
 F_2^A(x,Q^2)&=&A\int\frac{d\epsilon d^3p}{(2\pi)^4}\mathcal{P}_0(\epsilon,\mathbf{p})\frac{(1+\gamma\frac{p_z}{M})}{\gamma^2}\left( 1+4\frac{p^2x'^2}{Q^2}+6\frac{x'^2\mathbf{p}^2_\bot}{Q^2}\right) F_2^N\left( \frac{x}{z},Q^2\right),\label{eq:4}\\
 \fl
F_3^A(x,Q^2)&=&A\int\frac{d\epsilon d^3p}{(2\pi)^4}\mathcal{P}_0(\epsilon,\mathbf{p})\left( 1+\frac{p_z}{\gamma M}\right) \frac{x'}{x} F_3^N\left( \frac{x}{z},Q^2\right), \label{eq:5}
\end{eqnarray}
where $\mathcal{P}_0(\epsilon,\mathbf{p})$, which is the isoscalar part of the nuclear spectral function, is now normalized to unity and we perform the calculations for the nuclear structure functions averaged over neutrinos and antineutrinos, i.e., we only consider the symmetric $\nu+\bar{\nu}$ combination~\cite{Kulagin:2007ju,Kulagin:2004ie}.

In a nucleus, the expected values of $\frac{\left\langle \epsilon\right\rangle }{M}$ and $\frac{\left\langle \mathbf{p}^2\right\rangle }{M^2}$ averaged with the nuclear spectral function are much smaller than unity. Thus,  
$z \approx 1$ and $x'\approx x$. Under these assumptions, we can perform a Taylor expansion  of the integrands in expressions (\ref{eq:4}) and (\ref{eq:5}) around $z=1$, keeping terms up to order $\epsilon/M$ and $\mathbf{p}^2/M^2$. In this way we will be able to take out of the integral the structure functions and their derivatives and we will be left with expected values of the removal energy $\epsilon$ and momentum squared $\mathbf{p}^2$. This statement is true if the nuclear spectral functions only depend on the modulus of the momentum $|\mathbf{p}|$ (as it is in the case of the nuclear spectral functions we will consider) and not upon its direction. 
Under this assumption we can drop the expected values of $p_z$ (or any other component of the momentum) and 
 $\epsilon\, p_z$ because they are identically zero due to symmetry considerations.

We will begin with the structure function $F_2^A\left( x,Q^2\right) $. After performing the Taylor expansion, keeping terms up to order $\epsilon/M$ and $\mathbf{p}^2/M^2$, and dropping those terms which go with $\left\langle p_z\right\rangle /M$ or $\left\langle \epsilon\, p_z\right\rangle /M^2$ (because they are identically zero as stated above), we obtain 
\begin{eqnarray}
\fl
 \frac{F_2^A\left( x,Q^2\right) }{A}&\simeq&F_2^N\left( x,Q^2\right) \left[ 1+\left( \gamma^2-1\right) \frac{\left\langle \mathbf{p}^2\right\rangle }{3M^2}\right] -x\frac{\partial F_2^N\left( x,Q^2\right) }{\partial x}\left[ \frac{\left\langle \epsilon\right\rangle }{M}+\left( 2-\gamma^2\right)\frac{\left\langle \mathbf{p}^2\right\rangle }{3M^2} \right] \nonumber\\
 \fl
&+&\left( x\frac{\partial F_2^N\left( x,Q^2\right) }{\partial x}+\frac{x^2}{2}\frac{\partial^2F_2^N\left( x,Q^2\right) }{\partial x^2}\right) \gamma^2\frac{\left\langle \mathbf{p}^2\right\rangle }{3M^2}\label{eq:6},
\end{eqnarray}
where $\left\langle \mathcal{O}\right\rangle $ is the expected value of the operator $\mathcal{O}(\epsilon,\mathbf{p})$ averaged with the isoscalar spectral function $\mathcal{P}_0(\epsilon,\mathbf{p})$:
\begin{equation}
 \left\langle \mathcal{O}\right\rangle =\int\frac{d\epsilon\,d^3p}{(2\pi)^4}\:\mathcal{P}_0(\epsilon,\mathbf{p})\:\mathcal{O}(\epsilon,\mathbf{p})\label{eq:7}.
\end{equation}
We have also used the fact that, with a nuclear spectral function which depends only on the modulus of the momentum, the expected value of momentum squared is shared equally among every squared component, i.e:
$\left\langle p_i^2\right\rangle =\frac{\left\langle \mathbf{p}^2\right\rangle }{3}$.
In the Bjorken limit, ($\gamma\rightarrow1$), (\ref{eq:6}) coincides with the expansions used by
Frankfurt {\it et al.} \cite{Frankfurt:1985ui} and Ciofi degli Atti {\it et al.} \cite{CiofiDegliAtti:1989eg}.

If we do the same for the nuclear structure function $F_3^A\left( x,Q^2\right) $, we obtain
\begin{eqnarray}
\fl
 \frac{F_3^A\left( x,Q^2\right) }{A}&\simeq&F_3^N\left( x,Q^2\right)-\frac{\left\langle \epsilon\right\rangle }{M}\left\lbrace F_3^N\left( x,Q^2\right)+x\,\frac{\partial F_3^N\left( x,Q^2\right)}{\partial x}\right\rbrace \nonumber\\
 \fl
&+&\frac{\left\langle \mathbf{p}^2\right\rangle }{3M^2}\left\lbrace \left( \gamma^2-1\right)\left[ F_3^N
+x\,\frac{\partial F_3^N}{\partial x}\right]+\gamma^2\left[ x\,\frac{\partial F_3^N}{\partial x}+\frac{x^2}{2}\,\frac{\partial^2F_3^N}{\partial x^2}\right]   \right\rbrace \label{eq:9},
\end{eqnarray}
a similar result to that Kulagin~\cite{Kulagin:1997vv}. To allow for an easier comparison,
in  the above expressions, $\gamma^2$ can be rewritten as 
\begin{equation}
 \gamma^2=1+\frac{4M^2x^2}{Q^2}\label{eq:10}.
\end{equation}
Then, the only difference between (\ref{eq:9}) and that of \cite{Kulagin:1997vv} is that, for simplicity, we do not consider the off-shell dependence in the nucleon structure function $F_3^N$. In our calculation, the only source of off-shell dependence is through the nuclear spectral function $\mathcal{P}_0$. Thus, we do not have the term $\partial_{p^2}F_3^N$ that appears in \cite{Kulagin:1997vv} .

For our study  we have selected the $^{40}Ca$ nucleus, which is isoscalar and
already will show important medium effects. The free nucleon structure functions have been taken from \cite{Gluck:1998xa} and we have chosen several  different nuclear spectral functions. 
 The first one (labelled I)
is a phenomenological model of the spectral function which has a mean field part and 
high-momentum components coming from NN-correlations and it is described 
in  Kulagin {\it et al.}~\cite{Kulagin:2004ie} where it was used in a  global study of nuclear structure functions. 
The second one (labelled II), that also contains correlations is taken from
\cite{Ankowski:2007uy} where it has been tested in the calculation of several electron scattering observables. 
The third spectral function  is taken from the semiphenomenological model based on a many body calculation  
and the local density approximation
that is described in \cite{FernandezdeCordoba:1991wf} (labelled III). Finally, we also consider 
the simple mean field spectral function of \cite{Smith:2002ci}(labelled IV), which  was used in the study of the EMC effect.

\section{Results}
\label{sec:Re}

\begin{figure}
\includegraphics[width=0.9\textwidth]{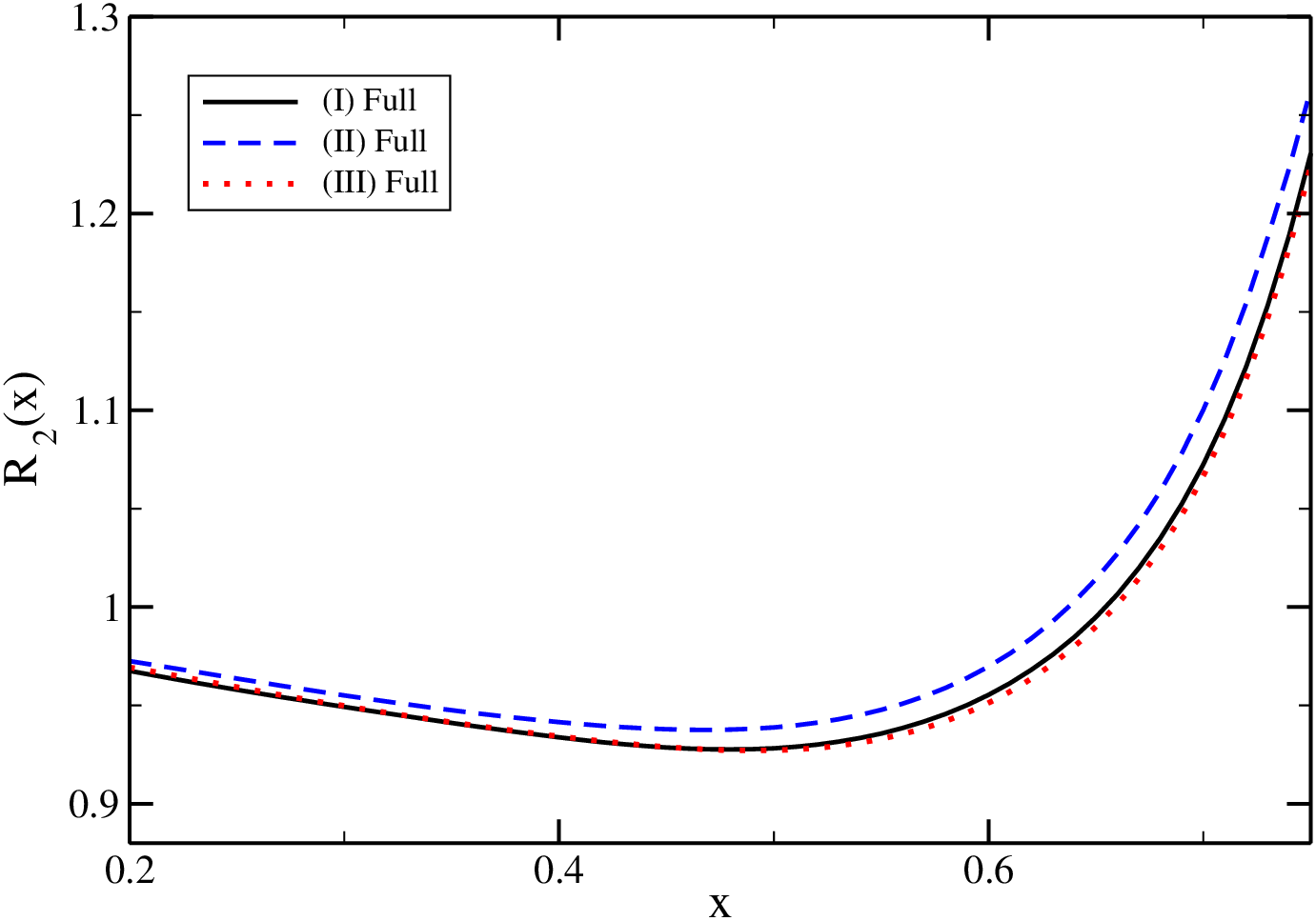}
\caption{$R_2=\frac{F_{2A}}{F_2}$ ratio for $^{40}Ca$ at
 $Q^2=20\;\textrm{GeV}^2$ with the nuclear spectral functions (I), (II) and (III) described in the text.}
\label{fig1}\end{figure}

In figure~\ref{fig1}, we show the results of the ratio $R_2=\frac{F_{2A}}{F2}$ for the nuclear spectral functions (I), (II) and (III). This gives us some idea of the uncertainties related to these functions. 
The differences are small, even when the spectral funtions have been  obtained with diverse methods and are in fact quite different if one studies in detail their energy and momentum dependence. However,
the expected values of the mean removal energy $\left\langle \epsilon\right\rangle $ and the mean kinetic energy per nucleon $\left\langle T\right\rangle =\left\langle \frac{\mathbf{p}^2}{2M}\right\rangle $ are quite similar, as can be seen in Table~\ref{tab:1}.
\begin{table}[h]
\caption{Expected values of the nucleon removal and kinetic energies for the nuclear spectral functions of Kulagin {\it et al.}~\cite{Kulagin:2004ie} (I),
Ankowski {\it et al.}~\cite{Ankowski:2007uy} (II) and Fernandez de Cordoba {\it et al.}~\cite{FernandezdeCordoba:1991wf} (III).}\label{tab:1}
\begin{indented}
\item[]\begin{tabular}{|l|c|c|c|}
\hline
Spectral Function & (I)& (II) & (III) \\ 
\hline
$\left\langle \epsilon\right\rangle$ (MeV) & -49 & -40 & -47 \\
\hline
$\left\langle T\right\rangle$ (MeV) & 30.1 & 26.2 & 28.8\\
\hline
\end{tabular}
\end{indented}
\end{table}
In particular, spectral functions (I) and (III) that have quite close expected values also produce very similar ratios,
whereas (II) which has an appreciably smaller binding energy gives slightly larger values for the ratio.

\begin{figure}
\includegraphics[width=0.9\textwidth]{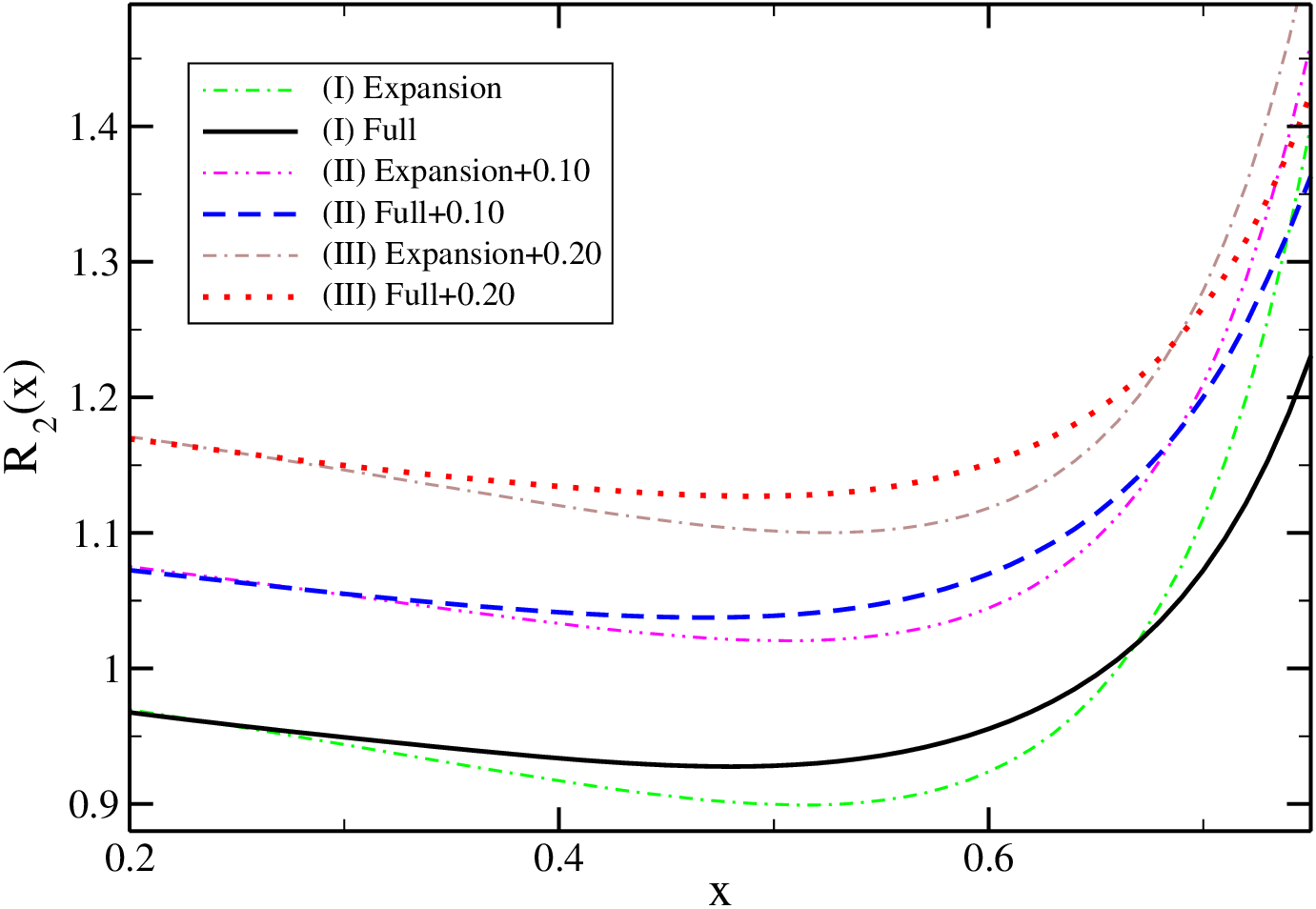}
\caption{$R_2=\frac{F_{2A}}{F2}$ ratio for $^{40}Ca$
at $Q^2=20\;\textrm{GeV}^2$. Comparison of the full results with the approximation of (\ref{eq:6}) for the nuclear spectral functions (I), (II) and (III).}\label{fig2}
\end{figure}

In figure~\ref{fig2}, we compare the full results for the same ratio (\ref{eq:4}) whith those obtained making use of the approximation of (\ref{eq:6}). The use of this expansion has been assumed to be a good approximation 
for $x\leq 0.5$~\cite{CiofiDegliAtti:1989eg,CiofidegliAtti:1990dh,CiofidegliAtti:1991ae}.
Obviously, the series expansion agrees well at low values of $x$ with the full results. However, in the three cases produces
lower values for the ratio at intermediate $x$ showing a maximum deviation of a 3-4$\%$ around $x=0.5-0.6$. 
This region, with a dip in the ratio, is dominated by the mean removal energy (or equivalently the binding energy) per nucleon. Although this could look a small error, we should remark that it means increasing the deviation due to nuclear effects from the value $1$ by around a 30\% .
At higher $x$'s, where Fermi motion of the nucleons provides the dominant effect, the series expansions grow faster than the full results and become larger for $x\gtrsim 0.65$. Thus, we find that for typical nuclear spectral functions
the convergence of the series expansion is not  so good except at very low $x$, where in any case other effects not considered here, like shadowing play a major role.
\begin{figure}
\includegraphics[width=0.9\textwidth]{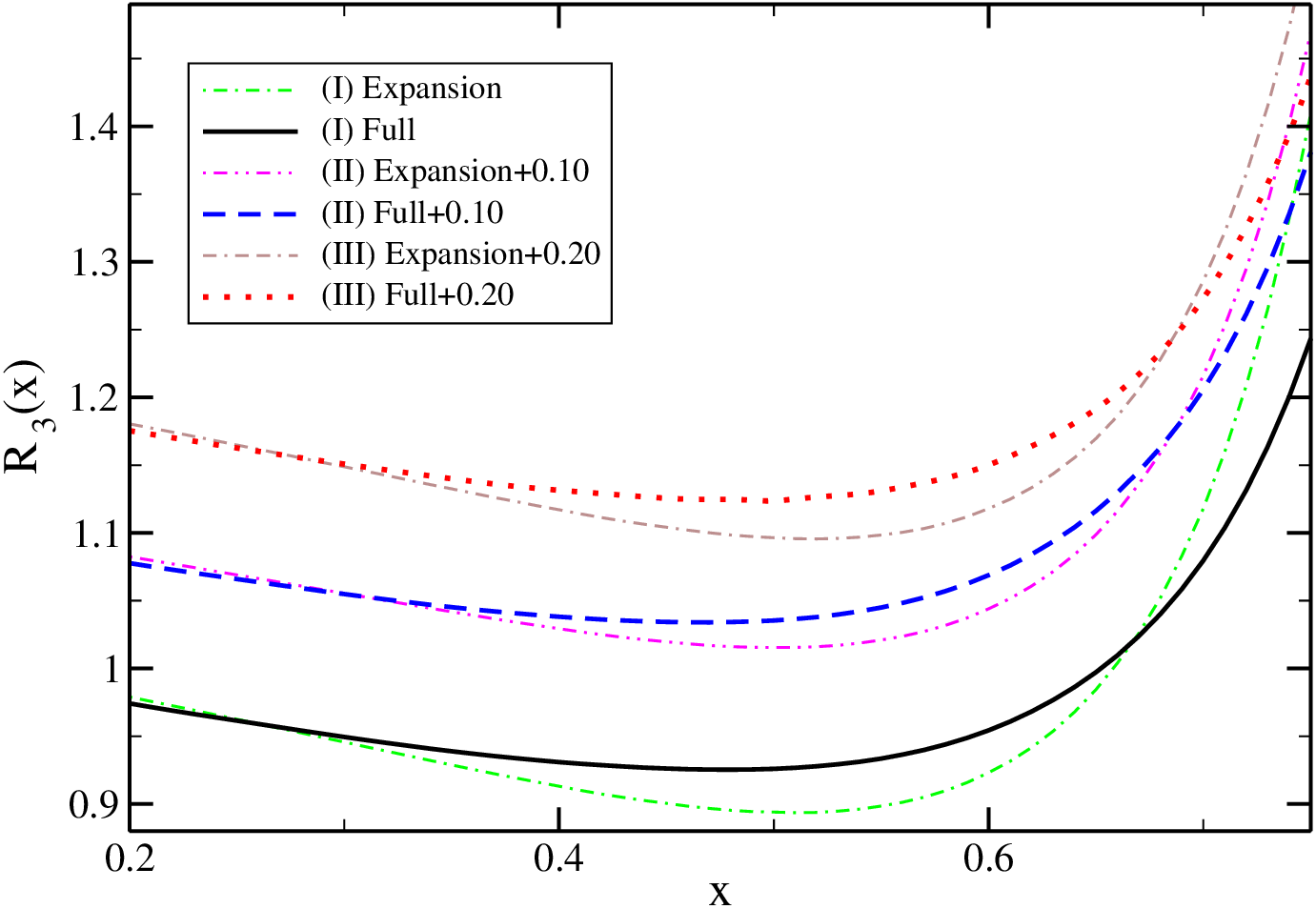}
\caption{$R_3=\frac{F_{3A}}{F_3}$ ratio for $^{40}Ca$
at $Q^2=20\;\textrm{GeV}^2$. Comparison of the full results with the approximation of (\ref{eq:9}) for the nuclear spectral functions (I), (II) and (III).}
\label{fig3}
\end{figure}

The results for the $R_3=\frac{F_{3A}}{F_3}$ ratio are shown in figure~\ref{fig3}. 
The full model for $R_3$ is given by (\ref{eq:5}) and the series expansion by (\ref{eq:9}).
As it was the case for $R_2$, there is a dip region dominated by the binding energy and the Fermi motion of the nucleons produces the large rising at high values of $x$.
The comparison of the full results and the series expansions shows the same features as for $R_2$. The expansions systematically underestimate the ratios at intermediate values of $x$ and overestimate them for $x>0.7$. 
This overestimation of the effect of the Fermi motion was already discussed in \cite{Kulagin:1997vv}.
There, it was also claimed that in the limit of high $Q^2$ and for heavy nuclei the expansion should be a good approxiamtion 
up to $x\lesssim0.75$. However, after studying $R_2$ and $R_3$ at different $Q^2$ values, apart from the one shown in
this paper, we have found that for medium nuclei the convergence of the series is only good at very low $x$ where other nuclear effects are very relevant. 

In order to obtain a better convergence one would need to reach a higher order in the expansions. This implies also 
the sensitivity to nuclear expected values of higher powers of the nucleons momenta as $<(p/M)^4>$. One should notice
that due to correlations these expected values are not negligible~\cite{CiofidegliAtti:1991mm,CiofidegliAtti:1995qe}. The three spectral functions considered above 
contain high momentum components and one may expect any expansion up to order $p^2$ to fail to have a good convergence.
To test this point, we have also calculated $R_2$ for the case of a simpler mean field spectral function which does not incorporate nucleon correlations and does not have those high momentum components~\cite{Smith:2002ci}. 
\begin{figure}
\includegraphics[width=0.9\textwidth]{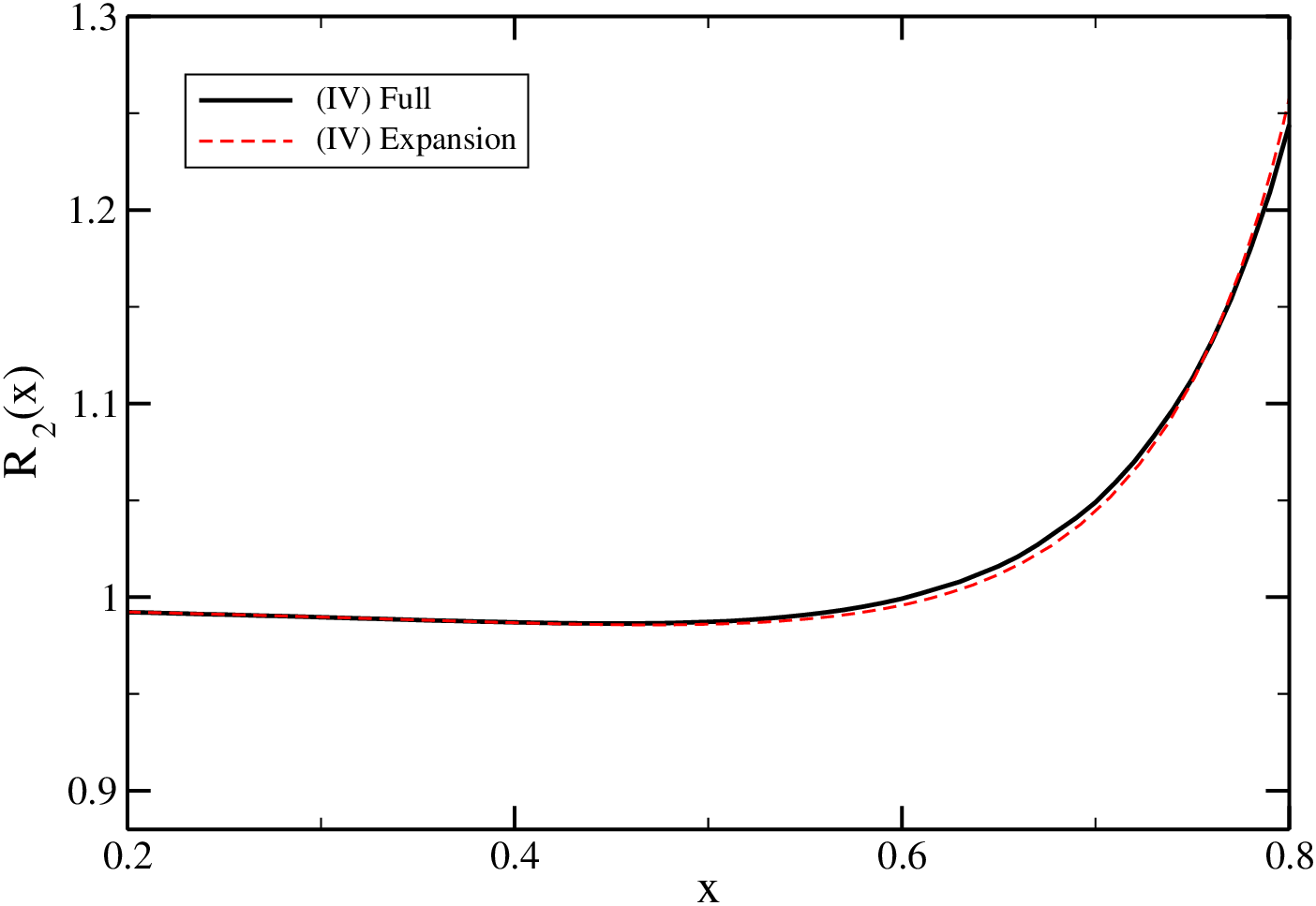}
\caption{$R_2=\frac{F_{2A}}{F2}$ ratio for $^{40}Ca$. Comparison of the full results with the approximation of (\ref{eq:6}) for the nuclear spectral functions (IV).}
\label{fig4}
\end{figure}
This function was used to analyse the A dependence of the position and magnitude of the dip.
Here,  we have used the same parton distribution functions as in the original reference~\cite{Smith:2002ci}.
The results are shown in figure~\ref{fig4}. In this case, there is an almost perfect agreement between the full results and the expansion.

In summary, we have studied the quality of some series expansions commonly used to incorporate approximately the nuclear effects in the analysis of DIS processes. We have found that for realistic enough nuclear  spectral functions, that include nucleon correlations and have high momentum components, the convergence of the series to the full result is poor except at very low values of $x$, where in fact, other nuclear effects, like shadowing or antishadowing,  are more relevant.
At high $x$ values, in the Fermi motion region, the expansions clearly overestimate the full result. This was known and expected. However, even at relatively low $x$ values, where the expansions were suppossed to provide a good approximation, we have found that they  systematically underestimate the value of $F_2$ and $F_3$, artificially increasing the size of the nuclear corrections.

\begin{ack}
This work was partially supported by the MEC contract
FIS2006-03438 and the EU Integrated Infrastructure
Initiative Hadron Physics Project contract RII3-CT-
2004-506078. I.R.S. acknowledges MEC for additional support.
\end{ack}


\end{document}